%% file: sample-acmtog-SIGGRAPH-submission.tex
\documentclass[acmtog]{acmart}
\renewcommand\footnotetextcopyrightpermission[1]{} 
\acmSubmissionID{1402}

\usepackage{booktabs} 

\usepackage{cleveref}
\input{preamble-sig}

\usepackage[ruled]{algorithm2e} 

\SetAlFnt{\small}
\SetAlCapFnt{\small}
\SetAlCapNameFnt{\small}
\SetAlCapHSkip{0pt}

\begin{document}
\title{\method: Feed-Forward Relightable Neural Gaussians}

\author{Guangming Fu$^\dagger$}
\thanks{$^\dagger$Equal contribution.}
\affiliation{
    \institution{Nankai University}
    \country{China}
}
\email{2120240685@mail.nankai.edu.cn}

\author{Jiahui Fan$^\dagger$}
\affiliation{
    \institution{Nanjing University}
    \country{China}
}
\email{jiahui.fan.1998@gmail.com}

\author{Jian Yang}
\affiliation{
    \institution{Nankai University}
    \country{China}
}
\email{csjyang@njust.edu.cn}

\author{Milo\v{s} Ha\v{s}an}
\affiliation{
    \institution{NVIDIA}
    \country{USA}
}
\email{milos.hasan@gmail.com}

\author{Beibei Wang$^\ddagger$}
\orcid{0000-0001-8943-8364}
\thanks{$^\ddagger$Corresponding author.}
\affiliation{
    \institution{Nanjing University}
    \country{China}
}
\email{beibei.wang@njust.edu.cn}

\renewcommand\shortauthors{Fu G. et al.}

\input{sec-sig/0_abstract}    


%
%

\ccsdesc[500]{Computing methodologies~Rendering}
%
%

\keywords{Relighting, Gaussian splatting, Feed-forward network, Neural appearance}

\begin{teaserfigure}
  \includegraphics[width=\textwidth]{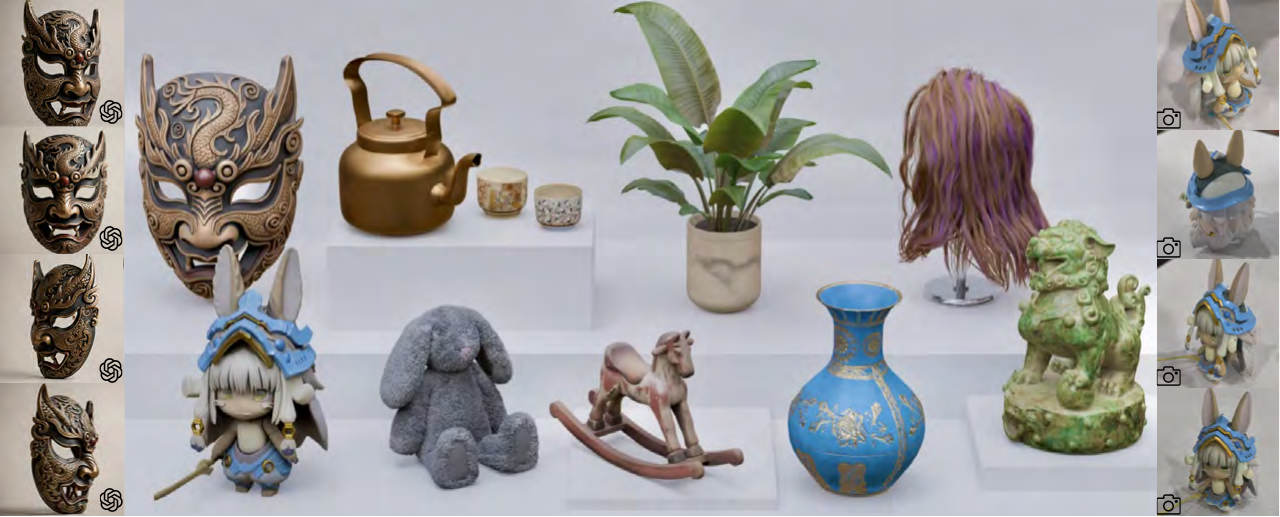}
  \caption{
   We propose \method, a novel feed-forward relighting framework that directly reconstructs relightable neural Gaussian assets from sparse view inputs. \method utilizes large model priors, while not re-training or fine-tuning any of them. The reconstructed assets can be rendered under any novel light or view conditions, supporting diverse open-category objects with complex appearances, such as fur. \method can also reconstruct and relight assets from generated multi-view images by GPT-Image-2~\cite{zewde2026gpt} (left), as well as photos captured from a real-world figurine (right).
  }
  \label{fig:teaser}
\end{teaserfigure}

\maketitle

\input{sec-sig/1_intro}
\input{sec-sig/2_related}

\input{sec-sig/3_method}

\input{sec-sig/4_results}
\input{sec-sig/5_discussion}


\clearpage

\bibliographystyle{ACM-Reference-Format}
\bibliography{sample-bibliography}

\input{sec-sig/sig_figure_only_pages}


\input{sec-sig/X_suppl}

\end{document}

%% file: preamble-sig.tex
\newcommand{\method}{F-RNG\xspace}

\usepackage{pifont}
\usepackage{graphicx}
\usepackage{subcaption}
\usepackage{multirow}
\usepackage{makecell, multicol, diagbox}
\usepackage{xcolor, colortbl}
\usepackage{array}
\newcolumntype{C}[1]{>{\centering\arraybackslash}p{#1}}

\newcommand{\cellwidth}{0.07}

\def\figurePath{fig/}
\def\myfigure#1#2#3{
\begin{figure}[tb]\centering\includegraphics[width = \linewidth]{\figurePath#2}\caption{#3}\label{fig:#1}
\end{figure}}
\def\mycfigure#1#2#3{
\begin{figure*}[t]\centering\includegraphics*[clip, width = \linewidth]{\figurePath#2}\caption{#3}\label{fig:#1}
\end{figure*}}

\usepackage{gensymb}

\definecolor{tabfirst}{rgb}{1, 0.7, 0.7} 
\definecolor{tabsecond}{rgb}{1, 0.85, 0.7} 
\definecolor{tabthird}{rgb}{1, 1, 0.7} 

%% file: sec-sig/0_abstract.tex
\begin{abstract}

Capturing relightable 3D assets from real-world objects is a widely researched problem. Several per-scene optimization-based methods, based on 3D Gaussian splatting (3DGS), support relighting; however, they usually require dense input views, and their overfitting nature makes it difficult to generalize across scenes. Unlike per-scene optimization methods, generalized feed-forward models can directly reconstruct Gaussians from sparse input views. However, the resulting assets have baked-in illumination and cannot be easily used for relighting.
In this paper, we present \method, a feed-forward framework that directly generates relightable 3DGS assets from sparse-view inputs. 
Training such a model from scratch can require massive data and computing resources, and it is especially challenging to generate relightable assets in a feed-forward manner with acceptable cost. To achieve this, we develop \method upon an existing large reconstruction model (LRM) to extract relightable representations, while also utilizing priors from an intrinsic decomposition model (IDM). 
Specifically, we aim to improve the precision of the underlying geometry and decompose the light and materials. We first introduce a latent-interpolated fine-grained geometry synthesis to enhance the LRM's geometry representation. Second, we propose a prior-guided relightable appearance distillation to extract relightable neural representations by incorporating IDM priors. Finally, a universal neural renderer enables flexible and high-fidelity relighting under arbitrary novel lights.
\method is a ``plug-and-play'' framework that requires neither re-training nor fine-tuning of the underlying LRMs, thus can automatically benefit from better LRMs and IDMs in the future. With only small networks that can be trained with affordable data and computational resources, \method directly generates relightable 3D assets, avoiding the repetitive inference of large models under different light conditions. By comparison to the state-of-the-art LRM-based relighting method, \method achieves $\sim$25$\times$ faster relighting, as well as superior quality ($\sim$+2.0 dB).
\end{abstract}

%% file: sec-sig/1_intro.tex
\section{Introduction}

Creating relightable 3D assets from multi-view images is an effective way to create digital content, with broad applications in entertainment, e-commerce, and virtual reality. However, due to the ambiguity in the decomposition of light, material, and geometry, the relighting problem still remains challenging, especially under unknown light conditions. 

Recent approaches, building upon 3D Gaussian splatting (3DGS) \cite{kerbl_2023_3dgs}, have shown impressive progress in decomposing the material, light, and geometry of objects. 
These optimization-based methods either introduce restrictive material or geometry priors~\cite{gao_2023_relightablegs, jiang_2024_gaussianshader, liang_2024_gsir}, or leverage controlled light conditions~\cite{bi2024rgs, bi_2020_neural} to achieve high-quality relighting. Recently, Fan et al.~\shortcite{fan2025rng} introduce Relightable Neural Gaussians (RNG), where the appearance of each Gaussian is represented by neural features (i.e., neural Gaussians), and is decoded via a per-scene overfitted multi-layer perceptron (MLP) decoder. Moreover, as the neural appearance modeling in RNG avoids extra geometry or shading model assumptions, it also supports many types of complex appearance, including fur. However, all methods above overfit to the training object, requiring dense views and/or controlled lighting conditions, thus usually take more than 30 minutes to converge for a single object. When creating assets from large-scale datasets, the time and computational cost of such a repetitive procedure are unacceptable.

Another line of work,  large reconstruction models (LRMs)~\cite{zhang2024gs, li2025lirm, siddiqui2024meta}, addresses the 3D reconstruction problem in a feed-forward fashion. They are typically trained on cross-scene 3D datasets, and predict scene representations from sparse input views directly. Specifically, RelitLRM~\cite{zhang2024relitlrm} trains a diffusion model that predicts radiances with baked-in illumination that is represented by spherical harmonics (SHs) under the target environment light. Although such diffusion-based models can generate visually plausible radiances, they fall short of producing \textit{true relightable 3D assets}: each new lighting requires a full inference pass of the large model, making the assets computationally heavy at runtime and difficult to insert into general 3D scenes.

In this paper, we propose \method, a novel feed-forward relighting framework that directly generates relightable neural Gaussian assets and performs flexible relighting through a generalized neural material representation. 
However, training such a feed-forward model from scratch usually requires massive data and computing resources, which are not easily accessible. To overcome this issue, we propose to utilize priors from existing large models, without re-training or fine-tuning them, but only add relatively small extra networks instead. 
Specifically, we choose an LRM, RelitLRM~\cite{zhang2024relitlrm}, to be our backbone, and we also introduce material priors from an intrinsic decomposition model (IDM), DiffusionRenderer~\cite{liang2025diffusion}.

The design of \method targets both geometry and appearance quality. First, we propose \textit{a latent-interpolated fine-grained geometry synthesis} to enhance the initial underlying geometry from the prior LRM. Second, we extract neural appearance representations from RelitLRM's radiance fields by a \textit{prior-guided relightable appearance distillation}, resulting in relightable neural Gaussian assets. In order to distill the appearance, we propose a Transformer network, named \textit{MaterialFormer}, to predict appearance features, as well as a \textit{light-independence regularization} to better eliminate the ambiguity between light and materials in a self-supervising manner. Finally, we render the relightable neural Gaussian assets by a \textit{universal neural renderer}, where the neural appearance modeling avoids assumptions on specific shading models and supports diverse complex appearances (e.g., fur). 
Integrating these key components enables \method  to reliably reconstruct from multi-view input, facilitating plausible decomposition and high-quality relighting while being compatible with common rendering pipelines.

We demonstrate the performance of \method by comparing with state-of-the-art relighting methods. Compared to RelitLRM, \method  can produce higher ($\sim$+2.0 dB) relighting quality across multiple real and synthetic datasets. 
In terms of computational efficiency, it takes us only 0.3 seconds to render with novel lighting or view ($512\times512$ resolution on an NVIDIA L40 GPU 48 GB), while RelitLRM~\cite{zhang2024relitlrm} needs more than 7 seconds ($\sim$25$\times$ slower). 
Moreover, \method is designed as a standalone enhancement framework that requires zero re-training or fine-tuning of the pre-trained LRMs or IDMs. This allows our method to directly benefit from future advancements in LRMs or IDMs.
Our contributions can be summarized as follows:
\begin{itemize}
\item  \textbf{\method, a novel feed-forward relighting framework} that directly generates relightable 3D assets from sparse-view inputs.
\item \textbf{a prior-guided relightable appearance distillation}, featuring a \textit{MaterialFormer} and \textit{light-independence regularization}, that leverages LRM and IDM priors to extract relightable neural representations, and
\item \textbf{a latent-interpolated fine-grained geometry synthesis} that produces more detailed underlying geometries and high-frequency details than the LRM backbone.
\end{itemize}

%% file: sec-sig/2_related.tex
\section{Related work}

\subsection{NeRF/3DGS variants for relighting}

Advancements in 3D reconstruction like neural radiance fields \cite{mildenhall_2020_nerf} and 3DGS~\cite{kerbl_2023_3dgs} achieve  Novel View Synthesis (NVS) with optimization-based pipelines, inspiring extensive research that pursues higher-quality appearances~\cite{barron_2021_mipnerf, barron_2022_mipnerf360, wang_2021_neus, chen_2023_neusg} and geometries~\cite{wang_2021_neus, dai_2024_gaussiansurfels, huang_2024_2dgs, guedon_2024_sugar}, or faster speed~\cite{chen_2022_tensorf}. Based upon them, several methods~\cite{zhang_2021_physg, liu_2023_nero, boss_2021_neuralpil} explicitly decompose the light, material, and geometry with multi-view images (often hundreds), achieving relighting under arbitrary novel lighting conditions. 

To achieve decomposition, several approaches \cite{bi_2020_neural, zeng2023relighting, bi2024rgs} rely on capturing data under restricted lighting conditions. 
Alternatively, many works \cite{jin2023tensoir, jiang_2024_gaussianshader, liang_2024_gsir} leverage analytical BRDFs as constraints to achieve physically-based decomposition, and explore inverse rendering to jointly optimize shape and material properties \cite{shi_2023_gir, Munkberg_2022_CVPR, sun2025svg, zhu2025gs, zhu2025gaussian}.
However, these analytical methods often struggle to generalize to complex appearances such as fur. Recently, Fan et al.~\shortcite{fan2025rng} presented RNG, which produces high-quality relightable representations for a wide range of object and material types. We employ a similar neural appearance modeling in our method.

At the core of all these methods is the per-scene optimization that iteratively solves for the light, material, and geometry factors. Thus all these methods overfit to the training scene and do not build any priors to generalize to unseen objects. Although the obtained relightable 3D assets can be rendered with arbitrary lights, the models still have to be optimized from scratch for every new object.  

\subsection{Material and shape reconstruction with LRMs}

LRMs \cite{hong2023lrm} and their variants~\cite{xu2024grm, zhang2024gs, ziwen2025long} enable feed-forward NVS from sparse-view inputs. Beyond them, some also explicitly focus on decomposing geometry and materials, making it more compatible with traditional rendering pipelines. For example, Wei et al.~\shortcite{wei2024meshlrm} and Xu et al.~\shortcite{xu2024instantmesh} propose to extract meshes and textures from reconstructed objects. 
Recently, AssetGen~\cite{siddiqui2024meta} and LIRM~\cite{li2025lirm} also propose to recover analytical texture maps in a feed-forward manner. However, compared to our relightable neural representations, the shading model and geometry assumptions in their methods also make it difficult to generalize to real-world objects with various appearances like fur and hair.

As our backbone, RelitLRM~\cite{zhang2024relitlrm} is the most related work to \method. With the target environment map as condition during the diffusion denoising process, RelitLRM directly predicts the radiance represented by SHs under the target illumination. However, it requires a full denoising pass for every new lighting, and the distant lighting assumption prevents easy integration in common rendering pipelines, since it does not support local or indirect bounce lighting.

\subsection{NVS with sparse-view inputs}


A typical 3DGS pipeline is often constructed based on dense inputs. Recently, this framework has been adapted for sparse views by several approaches~\cite{chung2024depth, zhu2024fsgs, li2024dngaussian, feng2024geogs3d}. Both approaches optimize discrete 3D Gaussians iteratively under multi-view supervision.  
A typical image-based reconstruction pipeline is data-intensive, and the ill-posed nature of the relighting problem makes it more difficult. Some methods target NVS with sparse input views via regularized optimization \cite{jain2021putting, kim2022infonerf, niemeyer2022regnerf}, or introduce spatial priors \cite{han2025sparserecon, younes2024sparsecraft, shi2024zerorf}, or other learned multi-view priors \cite{chen2021mvsnerf, wang2021ibrnet, huang2024neusurf, johari2022geonerf}.
However, these optimization-based strategies often suffer from slow convergence and limited quality due to complex regularization terms. Alternatively, generalized networks like PixelSplat \cite{charatan2024pixelsplat} and MVSplat \cite{chen2024mvsplat} attempt to predict Gaussian parameters directly via feed-forward architectures. Although recent works such as DepthSplat \cite{xu2025depthsplat} and GSRecon \cite{yang2025gsrecon} have improved efficiency, 
these generalizable methods still experience degraded rendering quality compared to scene-specific optimization.

\subsection{Image-based relighting}

Single-image relighting is a challenging task, as the input cannot provide sufficient information about the 3D structure. 
Early methods~\cite{ren2015image, xu2018deep} utilize deep convolutional networks to synthesize novel light conditions of the target image. Subsequent research \cite{sun2019single, zhou2019deep, pandey2021total, mei2024holo} focused heavily on portrait relighting to improve quality and realism.
Another line of research \cite{bell2014intrinsic, kocsis2024intrinsic, careaga2023intrinsic} focuses on the intrinsic decomposition (i.e., the IDMs).
With the advent of diffusion models, direct relighting \cite{zeng2024dilightnet, poirier2024diffusion, jin2024neural, kocsis2024lightit, zhang2025scaling} and intrinsic decomposition \cite{luo2024intrinsicdiffusion, zeng2024RGBX, he2026unirelight} have revolutionized. 
Most recently, Litman et al.\shortcite{litman2025lightswitch} introduced LightSwitch, which incorporates the material prediction diffusion model from MaterialFusion \cite{litman2025materialfusion} to condition a relighting U-Net. In a parallel direction, Liang et al.\shortcite{liang2025diffusion} proposed DiffusionRenderer, an IDM that adapts video diffusion models to predict shading buffers as a relightable representation. 
In \method, we introduce DiffusionRenderer as our prior IDM for an instance. \method is not limited to any IDMs, but is compatible and can benefit from any advanced models in the future.


%% file: sec-sig/3_method.tex
\mycfigure{pipeline}{pipeline-0.pdf}{
\textbf{The overview of \method.} 
\method leverages priors from an LRM and IDM, and it consists of three main components: (A) a \textit{latent-interpolated fine-grained geometry synthesis} that produces detailed geometries, (B) a \textit{prior-guided relightable appearance distillation}, comprising of \textit{MaterialFormer} and a \textit{light-independence regularization}, to extract relightable neural representations, and (C) a \textit{universal neural renderer} that can render neural Gaussians under arbitrary lights via an MLP decoder.
}

\section{Preliminaries}

\method uses RelitLRM~\cite{zhang2024relitlrm} as its backbone, and utilizes priors from DiffusionRenderer~\cite{liang2025diffusion} to efficiently extract relightable 3D assets from input images. In this section, we briefly introduce these two prior techniques.

\subsection{RelitLRM}

RelitLRM~\cite{zhang2024relitlrm} reconstructs 3D Gaussian objects from sparse views $I_i$ and Pl\"{u}cker rays $\mathbf{R}_i$. It employs a geometry transformer to predict spatial Gaussian attributes, and an appearance generator, which is a Diffusion Transformer (DiT), to synthesize view-dependent radiance that is represented by SHs. The pipeline is formulated as:
\begin{align}
    \{ \mathbf{T}_{i}^{\text{geo}} \} &= \operatorname{GeoTransformer}(\operatorname{Patchify}(I_{i}, \mathbf{R}_{i})), \\
    \{ \mathbf{G}_{i}^{\text{geo}} \} &= \operatorname{GeoDetokenizer}( \{\mathbf{T}_{i}^{\text{geo}} \}), \\
    \{\mathbf{T}_{i}^{\text{all}}\} &= \left\{ \{\mathbf{T}_{i}^{\text{noised}}\} \oplus \{\mathbf{T}_{i}^{\text{light}}\} \oplus \{\mathbf{T}_{i}^{\text{geo}} \} \oplus \{\mathbf{T}_{i}^{\text{step}}\} \right\}, \\
    \{ \mathbf{G}_{i}^{\text{app}} \} &= \operatorname{AppDetokenizer}(\operatorname{DiT}(\{\mathbf{T}_{i}^{\text{all}} \})),
\end{align}
where all the de-tokenizers are single-layer MLPs. However, RelitLRM falls short of producing \textit{true relightable 3D assets}, as it must run the diffusion model repeatedly for every novel lighting condition, producing a new non-relightable asset every time and incurring substantial time and computational overhead. 

\subsection{DiffusionRenderer}

Intrinsic decomposition models typically take an image or video as input to estimate per-pixel properties such as albedo, lighting, depth, normals, roughness, or other material parameters. We use the recent DiffusionRenderer model~\cite{liang2025diffusion}, specifically its inverse rendering component. 
In DiffusionRenderer, an input video $V$ is encoded as a condition into a latent vector $z=\mathcal{E}(V)$, and an intrinsic attribute $P$ is generated by a fine-tuned diffusion model. Repeating this process for all attributes $P$ yields a shading buffer set $\{\mathbf{\hat{g}}_{0}^{n}, \mathbf{\hat{g}}_{0}^{d}, \mathbf{\hat{g}}_{0}^{a}, \mathbf{\mathbf{\hat{g}}}_{0}^{r}, \mathbf{\hat{g}}_{0}^{m}\}$, which refers to the normal, depth, albedo, roughness and metallic parameters of the Cook-Torrance shading model~\cite{cook1982reflectance}. These G-buffers, namely IDM priors, are then used as guidance in the following modules of \method.
Here, we only introduce the IDM prior as a guidance, instead of as a direct supervision, as the intrinsic decomposition can be undefined or less accurate for complex appearances (e.g., fur or sub-surface scattering). However, such a guidance can still provide us with plausible lighting removal, and thus help to  predict our neural reflectance model.

\section{Methods}


\method considers both geometry and appearance quality; its pipeline is shown in Fig.~\ref{fig:pipeline}.
First, \method performs a \textit{latent-interpolated fine-grained geometry synthesis} (Sec.~\ref{sec:densification}) to generate more detailed underlying geometries. Second, \method designs a \textit{prior-guided relightable appearance distillation} (Sec.~\ref{sec:appearance}) to achieve the decomposition, comprising two key components: the \textit{MaterialFormer} (Sec.~\ref{sec:delight}) that extracts relightable appearance representations from RelitLRM's radiance outputs, and the \textit{light-independence regularization} (Sec.~\ref{sec:independent}) that is designed to regularize the light/material decomposition. Finally, \method uses a \textit{universal neural renderer} (Sec.~\ref{sec:relightable}) to achieve high-quality rendering under arbitrary lights.

\subsection{Latent-interpolated fine-grained geometry synthesis}
\label{sec:densification}

\myfigure{geo_pipeline}{pipeline-1.pdf}{
\textbf{The overview of latent-interpolated fine-grained geometry synthesis.}
By detecting the top-$K$ salient patches with highly-detailed geometries or textures, \method interpolates RelitLRM's geometry tokens to sythesize new ones.  They are then decoded by another de-tokenizer to represent detailed structures, with the IDM priors as guidance.
\vspace{-7pt}
}

Although RelitLRM~\cite{zhang2024relitlrm} can produce reliable Gaussian geometries from input images, we still observe the loss in complex textures and high-frequency geometric details. This is primarily due to the intrinsic limitations of RelitLRM's per-pixel reconstruction paradigm, which limits the number and density of resulting Gaussians.
To address this issue, we propose the latent-interpolated fine-grained geometry synthesis. In addition to the original (coarse-grained) Gaussian geometry outputs from RelitLRM, we interpolate within the latent space for fine-grained Gaussians. Since the geometry decoder in RelitLRM is pre-trained at the original resolution, they can hardly interpret these new fine-grained Gaussians into more detailed geometries. Therefore, we propose a dedicated de-tokenizer for them. For the original geometry tokens from RelitLRM (i.e., coarse-grained ones), we still keep the pre-trained geometry decoder.

As shown in Fig.~\ref{fig:geo_pipeline}, the synthesis follows three steps: 1) First, we compute a saliency map to detect patches that need refinement; 2) Second, we select top-$K$ salient patches and interpolate their tokens to generate fine-grained ones; 3) At last, a fine-grained de-tokenizer is trained to decode these fine-grained tokens into Gaussians, guided by the IDM priors.
The whole process can be formulated as:
\begin{align}
 \{\textbf{T}^\mathrm{prior}_i\} &= \mathrm{Tokenizer} ( \mathrm{Patchify} ( \{\mathbf{\hat{g}}_{i}^{n}, \mathbf{\hat{g}}_{i}^{a}, \mathbf{\hat{g}}_{i}^{r}, \mathbf{\hat{g}}_{i}^{m} \})),\\
    \{\mathbf{T}_{j}^{\mathrm{fine-geo}}\} &= \operatorname{Bilinear}(\{\mathbf{T}_{j}^{\mathrm{geo}}\}), \\
    \{\mathbf{T}_{j}^{\mathrm{fine-prior}}\} &= \operatorname{Bilinear}(\{\mathbf{T}_{j}^{\mathrm{prior}}\}), \\
 \{  \mathbf{G}_{j}^{\mathrm{fine-geo}} \} = &  \operatorname{FineGeoDetokenizer}(\{\mathbf{T}_{j}^{\mathrm{fine-geo}}\} \mid \{\mathbf{T}^\mathrm{fine-prior}_{j}\}),
\end{align}
where both the tokenizer and de-tokenizer are trainable single-layer MLPs, and $j$ denotes the indices of the patches with the top responses in the saliency map.
By interpolating ray directions within each pixel's footprint, the geometry decoder can reconstruct finer-grained Gaussians at sub-pixel levels, compared to the original per-pixel Gaussians paradigm. 
The saliency map is designed to reflect the textural and geometric complexity while avoiding boundary effects, and the detailed algorithm for its computation is described in our supplementary. 

\subsection{Prior-guided relightable appearance distillation}
\label{sec:appearance}

\myfigure{app_pipeline}{pipeline-2.pdf}{
\textbf{The overview of prior-guided relightable appearance distillation.} 
This module consists of two components: the MaterialFormer and the light-independence regularization. 
RelitLRM (omitted in the figure) generates geometry and appearance tokens from input views; they are fed into MaterialFormer, together with the encoded IDM priors. The MaterialFormer outputs material tokens that are decoded to relightable appearance features, and these tokens are regularized by an extra loss term to be similar in spite of given different light conditions as inputs to RelitLRM.
}

While RelitLRM~\cite{zhang2024relitlrm} produces high-quality radiance fields, the resulting SH representations suffer from baked-in illumination, which limits their direct application in downstream relighting tasks.
In order to extract relightable representations from RelitLRM's predictions, we propose the prior-guided relightable appearance distillation. As shown in Fig.~\ref{fig:app_pipeline}, the distillation comprises two key components: the \textit{MaterialFormer }and the \textit{light-independence regularization}.
First, we propose MaterialFormer, a relatively small Transformer network to directly extract relightable neural representations from RelitLRM's outputs. Second, to better eliminate the ambiguity between light and materials, we propose the light-independence regularization for more constraints.

\subsubsection{MaterialFormer}
\label{sec:delight}

MaterialFormer is guided by IDM priors from DiffusionRenderer~\cite{liang2025diffusion}. First, all these IDM priors are patchified and encoded into IDM prior tokens. Here, we use the same tokenizer for IDM priors as in Sec.~\ref{sec:densification}. Then, the IDM prior tokens and the RelitLRM's geometry/appearance tokens are together fed into our MaterialFormer, which can be formulated as:
\begin{align}
    \{ \textbf{T}^\mathrm{all}_i \} & = 
    \left\{ \{\textbf{T}^\mathrm{geo}_i\} \oplus \{\textbf{T}^\mathrm{app}_i\} \right\}
    ,
    \\
    \{\mathbf{T}^{\mathrm{mat}}_i \} &=  \mathrm{MaterialFormer}( \{ \textbf{T}^\mathrm{all}_i \} \mid \{\textbf{T}^\mathrm{prior}_i\} ),
    \\
    \{ \mathbf{G}^\mathrm{mat}_i \} &= \mathrm{MatDetokenizer}(\{ \textbf{T}^\mathrm{mat}_i \} ),
\end{align}
where $\{ \mathbf{G}^\mathrm{mat}_i \}$ represents the neural material features for each predicted Gaussian, and the de-tokenizer is a single-layer MLP.
More details about the structure and optimization of MaterialFormer is described in our supplementary.
As reflectance attributes are naturally insensitive to spatial resolution, the input tokens of MaterialFormer can be either coarse-grained or fine-grained ones, due to the latent-interpolated fine-grained geometry synthesis. In that case, the appearance tokens are interpolated the same way as the geometry tokens. Here, we omit the superscript for simplicity. 

\subsubsection{Light-independence regularization}
\label{sec:independent}

A key insight is that material properties are intrinsic physical attributes of an object and should, by definition, be independent of extrinsic illumination. Consequently, for the same object, although RelitLRM predicts multiple radiance fields under varying light conditions, the MaterialFormer should exhibit output invariance. With this consistency, we effectively constrain the network to distill material tokens that are agnostic to input light conditions.

We enforce the light-independence regularization in a self-super\-vised manner. For each training sample, we provide RelitLRM~\cite{zhang2024relitlrm} with an extra random environment light $E_1$, which leads to the same geometry tokens $\{\mathbf{T}^\mathrm{geo}_{i}\}$ but different appearance tokens $\{ \mathbf{T}^\mathrm{app}_{i1} \}$. After processing through the MaterialFormer, the corresponding material tokens $\{\mathbf{T}^\mathrm{mat}_{i1}\}$ are generated. 
These extra tokens, together with the original one $\{\mathbf{T}^\mathrm{mat}_{i}\}$, are utilized to enforce the light-independence regularization, encouraging similar material predictions regardless of the varying illumination inputs. 
The light-independence loss is defined as follows:
\begin{equation}
    \mathcal{L}_\mathrm{ind} = \lambda_c\mathcal{C}(\{\mathbf{T}^\mathrm{mat}_{i}\}, \{\mathbf{T}^\mathrm{mat}_{i1}) \} +  \lambda_\mathrm{KLD}\mathcal{L}_\mathrm{KLD}(\{ \mathbf{T}^\mathrm{mat}_{i} \}, \{ \mathbf{T}^\mathrm{mat}_{i1} \} ),
\end{equation}
where $\mathcal{C}$ denotes the cosine similarity, and $L_\mathrm{KLD}$ denotes the Kullback-Leibler divergence ~\cite{kullback1951information}, which can better preserve the distributions of the learned high-dimensional latent space than the na\"{i}ve $\mathcal{L}_1$/$\mathcal{L}_2$ loss.


\subsection{Universal neural renderer}
\label{sec:relightable}

After extracting 3D Gaussian assets, the next step is to render it. 
Instead of physically-based shading models~\cite{cook1982reflectance}, we follow RNG~\cite{fan2025rng} to employ neural reflectance modeling in \method to support a wider range of object and material types (e.g., fur).
However, utilizing neural appearance representation in a generalized feed-forward framework is non-trivial. RNG trains a per-scene optimized MLP decoder to render neural features at each Gaussian. In our case, simply adopting this scene-specific solution as-is is impractical, as predicting the weights of an overfitted MLP directly from a large model is notoriously difficult to train. Instead, we generalize to a single universal neural appearance decoder for multiple objects across different categories.

In \method, each Gaussian is assigned a neural feature vector that models its reflectance, and we train a universal neural appearance decoder to make up the latent space for reflectance representation. The reflectance is represented as
\begin{equation}
\label{eq:neural}
    \rho( \{ \mathbf{G}_{i}^\mathrm{mat} \}, \mathbf{\omega}_o, \mathbf{\omega}_i) = \Theta( \{ \mathbf{G}_{i}^\mathrm{mat} \} \mid \mathbf{\omega}_o, \mathbf{\omega}_i),
\end{equation}
where $(\mathbf{\omega_o}, \mathbf{\omega_i})$ are the light/view directions in the object frame of reference, and $\Theta$ is an MLP trained across all objects in the dataset.  
We explain some designs of our decoder, and describe its detailed implementation, including the rendering process in our supplementary.

%% file: sec-sig/4_results.tex

\section{Results}

In this section, we first introduce the experiment setups for validation, then provide comparison and analysis of \method on varying datasets. 
In our supplementary, we also describe the data curation and model details for the reader's reference.

\subsection{Experiment setups}

We validate our method on diverse objects across various synthetic and real-world datasets. For the synthetic data, we demonstrate assets from \textsc{Objaverse}~\cite{deitke2023Objaverse}, \textsc{TensoIR}~\cite{jin_2023_tensoir}, \textsc{Rna}~\cite{mullia2024rna} and \textsc{Rng}~\cite{fan2025rng}. For the real data, we validate on \textsc{Stanford-Orb}~\cite{kuang2023stanfordorb} (\textsc{Orb}), \textsc{Objects-with-lighting} ~\cite{Ummenhofer2024OWL} (\textsc{Owl}). All results are rendered with 128 spp and a resolution of 512 $\times$ 512. Our method supports various number of input views, and following RelitLRM's best practice, we choose 6 input views in all our experiments for a fair comparison. We compute the peak signal-to-noise ratio (PSNR), structural similarity index measure (SSIM)~\cite{wang_2004_ssim}, and perceptual similarity (LPIPS)~\cite{zhang_2018_lpips} values for comparison, to evaluate both pixel-wise error and perceptual differences.

We choose the state-of-the-art relighting methods for the relighting quality comparison. 
Unfortunately, for LRMs, since the codes and models of LIRM~\cite{li2025lirm} and AssetGen~\cite{siddiqui2024meta} are not available, we only compare with our backbone, RelitLRM, on both real and synthetic datasets. 
Apart from feed-forward methods, we also compare with dense-view optimization-based 3DGS relighting methods (GS-IR~\cite{liang_2024_gsir} and R3DG~\cite{gao_2023_relightablegs}) on synthetic datasets to further validate our relighting quality.
Finally, we compare with recent image-based relighting methods (Neural Gaffer\cite{jin2024neural} and DiLightNet\cite{zeng2024dilightnet}) on synthetic datasets.

\subsection{Quality validation}
\label{sec:validation}

\input{tables/vs_lrm_syn+real}

\paragraph{Comparison with RelitLRM}
In Figs.~\ref{fig:vs_lrm_syn} and \ref{fig:vs_lrm_real} and Table~\ref{tab:vs_lrm_syn+real}, we compare the relighting quality of \method with RelitLRM~\cite{zhang2024relitlrm} on synthetic and real-world datasets. \method achieves superior quantitative metrics and visual quality, particularly for challenging complex materials like fur.
Crucially, our method achieves much faster inference ($\sim$25$\times$ speedup) during relighting compared to RelitLRM.
More visual results are provided in our supplementary.

\input{tables/tbl_exp_tensoir}

\paragraph{Comparison with optimization-based 3DGS relighting methods}
In Table~\ref{tab:tensoir_comp}, we compare our relighting results with dense-view 3DGS-based relighting approaches, GS-IR~\cite{liang_2024_gsir} and R3DG~\cite{gao_2023_relightablegs}. \method demonstrates superior quality using only 6 sparse input views. In contrast, those per-scene optimization approaches typically require over 100 dense input images and up to 30 minutes of optimization per object.
We also provide visual comparisons in our supplementary.

\begin{table}[t]
    \fontsize{8pt}{10pt}\selectfont
    \centering
    \caption{ \textbf{Comparison of our relighting quality with image-based relighting methods on \textsc{Objaverse} dataset}. 
    \method takes 6 views as input to reconstruct and relight the target unseen views, and other methods process each view individually to generate relighting results.
    The best/second-best results are marked in \colorbox{tabfirst}{red}/\colorbox{tabsecond}{yellow}.
    Please refer to Fig.~\ref{fig:vs_image_based} for visual comparison.
    }
    \vspace{-7pt} 
    \begin{tabular}{l|cccc}
    \toprule
        Methods  &   PSNR & SSIM & LPIPS   \\
      \midrule
      Ours  & \cellcolor{tabfirst}{31.77}  & \cellcolor{tabfirst}{0.94} & \cellcolor{tabfirst}{0.047}   \\
      Neural Gaffer & \cellcolor{tabsecond}{26.55}  & \cellcolor{tabsecond}{0.90} & \cellcolor{tabsecond}{0.092}   \\
      DiLightNet  & 20.34  & 0.84 & 0.175   \\
    \bottomrule
    \end{tabular}
    \label{tab:vs_img_based}
\end{table}

\paragraph{Comparison with image-based relighting methods}
In Fig.~\ref{fig:vs_image_based} and Table~\ref{tab:vs_img_based}, we compare the relighting quality with image-based relighting methods (Neural Gaffer \cite{jin2024neural} and  DiLightNet \cite{zeng2024dilightnet}) on \textsc{Objaverse}~\cite{deitke2023Objaverse} dataset.
To compare as fairly as possible, for our results, we reconstruct the object from 6 input views and relight under the target unseen view; for other methods, we directly relight each image individually.
While image-based relighting approaches leverage powerful image-space diffusion priors, they lack accuracy in decomposition and consistency across views without an underlying 3D representation.
Extra visual comparisons and quantitative metrics are also provided in our supplementary.

\paragraph{Relighting with multiple applications.}
In Fig.~\ref{fig:teaser}, we show a relighting result by \method with both synthetic and real objects. 
Among them, the Chinese dragon mask is first generated by a text-to-image model, GPT-Image-2~\cite{zewde2026gpt}, with prompt ``Chinese dragon mask, intricate carvings, red gem on forehead, dark fantasy style'', and then reconstructed and relit with \method. The figurine is captured by the authors, utilizing 6 unstructured images under unknown environmental lighting. Some of the input images are shown alongside.
With our framework, one can easily synthesize, capture and reconstruct relightable assets that can be used in multiple downstream tasks for flexible relighting with common rendering pipelines.


\paragraph{Additional validation.}
In our supplementary, apart from extra visual results of experiments above, we also compare our relighting quality with two additional image-based relighting methods (DiffusionRenderer and LightSwitch~\cite{litman2025lightswitch}) on a real-world dataset, compare our NVS quality with a sparse-view 3DGS method (DNG~\cite{li2024dngaussian}), and analyze the impact of the number of input views. Please refer to them for more information.


\subsection{Ablation study}
\label{sec:ablation}

We provide visual and quantitative comparisons for all main components of \method in Fig.~\ref{fig:ablation} and Table~\ref{tab:ablation} on \textsc{Objaverse} dataset.

\paragraph{Latent-interpolated fine-grained geometry synthesis.}
In \method, we perform latent-space interpolation to enhance the geometry representation. The generated fine-grained Gaussians can produce sharper appearances, especially for high-frequency details.

\paragraph{IDM priors.}
Without the IDM priors, our appearance distillation fails to achieve a plausible decomposition of light and materials, leading to biased colors.

\paragraph{Light-independence regularization.}
The light-independence regularization helps predict plausible material tokens. Without it, the relighting result can suffer from severe baked-in illumination.
In our supplementary, we also show the effect of the light-independence regularization by analyzing the latent-space distribution of learned features, and validate the robustness of our decomposition against different input light conditions.

\paragraph{Neural appearance modeling.}
Our neural reflectance achieves superior fidelity, especially in reproducing complex material appearances like hair and fur, offering a more flexible and accurate framework for open-category objects.


\input{tables/tbl_exp_ab1}

\myfigure{ablation}{ablation_all.pdf}{
\textbf{Ablation study.} 
(A) \textit{Latent-interpolated fine-grained geometry synthesis} overcomes resolution limits on high-frequency details, resulting in sharp textures in salient regions.
(B) \textit{IDM priors} provide guidance for the decomposition; without them, the network exhibits outputs with biased colors.
(C) \textit{Light-independence regularization} helps eliminate the ambiguity between light and material, effectively preventing the baked-in illumination.
(D) \textit{Neural appearance modeling} allows for complex appearances (such as fur), showing a closer match to the reference compared to the physically-based shading model.
}


%% file: tables/vs_lrm_syn+real.tex
\renewcommand{\cellwidth}{0.035}
\begin{table}[tb]
    \centering
    \caption{
    \textbf{  
    Comparison of our relighting quality with RelitLRM~\cite{zhang2024relitlrm}.}
    \method produces overall higher quality than RelitLRM on multiple synthetic and real-world datasets. Our method also shows significant improvement in the relighting speed ($\sim$25$\times$).
    The best results are marked in \colorbox{tabfirst}{red}.
    Please refer to Figs.~\ref{fig:vs_lrm_syn} and \ref{fig:vs_lrm_real} for visual comparison.
    }
    \label{tab:vs_lrm_syn+real}
    \vspace{-7pt} 
    \resizebox{0.44\textwidth}{!}
    {
    \begin{tabular}
    {l|ccc|ccc }
        \toprule
        
        Dataset & \multicolumn{3}{c|}{Ours} & \multicolumn{3}{c}{RelitLRM} \\
         & PSNR & SSIM & LPIPS & PSNR & SSIM & LPIPS \\
        
        \midrule

        \textsc{BlenderKit} 
        & \cellcolor{tabfirst}{30.02}& \cellcolor{tabfirst}{0.910} & \cellcolor{tabfirst}{0.083}
        & 27.63& 0.834 & 0.123
        \\
        \textsc{TensoIR} 
        & \cellcolor{tabfirst}{30.96} & \cellcolor{tabfirst}{0.948} & \cellcolor{tabfirst}{0.057} 
        & 28.66& 0.936 & 0.075
        \\
        \textsc{Objaverse} 
        & \cellcolor{tabfirst}{31.77} & {0.945} & \cellcolor{tabfirst}{0.047}
        & 30.02 & \cellcolor{tabfirst}{0.950} &  0.049
        \\
        \textsc{Stanford-Orb} 
        & \cellcolor{tabfirst}{32.24}& \cellcolor{tabfirst}{0.968} & \cellcolor{tabfirst}{0.028}
        & 30.72&  0.965 & 0.041
        \\
        \textsc{Owl}
        & \cellcolor{tabfirst}{26.03}& \cellcolor{tabfirst}{0.841} & \cellcolor{tabfirst}{0.244}
        & 23.14& 0.790 & 0.284
        \\
\midrule

    \textit{Relighting time} & \multicolumn{3}{c|}{\cellcolor{tabfirst}{246} ms} & \multicolumn{3}{c}{ 6,826 ms} 
\\
        \bottomrule
    \end{tabular}

    }
\end{table}

%% file: tables/tbl_exp_tensoir.tex
\newcommand{\bw}{0.35cm}
\newcommand{\bww}{0.4cm}
\newcolumntype{C}[1]{>{\centering\arraybackslash}p{#1}}
\begin{table}[tb]
    \fontsize{7.5pt}{8pt}\selectfont
    \centering
    \caption{
    \textbf{Comparison of our relighting quality with dense-view 3DGS relighting methods~\cite{liang_2024_gsir, gao_2023_relightablegs}.}
    While these optimization-based methods rely on over 100 views as input, \method uses only 6 images, delivering higher relighting quality. 
    We report PSNR, SSIM and LPIPS, and the best/second-best results are marked in \colorbox{tabfirst}{red}/\colorbox{tabsecond}{yellow}. 
    The visual comparison are also provided in our supplementary.
    }
    \label{tab:tensoir_comp}
    \vspace{-7pt} 
    {
    \begin{tabular}{l|C{\bw}C{\bww}C{\bww}|C{\bw}C{\bww}C{\bww}|C{\bw}C{\bww}C{\bww}}
        \toprule
        Scene & \multicolumn{3}{c|}{Ours\textit{ (6 views)}}  & \multicolumn{3}{c|}{GS-IR \textit{(100 views)}} & \multicolumn{3}{c}{R3DG \textit{(100 views)}} \\
        \midrule

        Armadillo & \cellcolor{tabfirst}34.94 &  \cellcolor{tabfirst}0.983& \cellcolor{tabfirst}0.028 & 27.65 & 0.908 &0.082 & \cellcolor{tabsecond}30.76& \cellcolor{tabsecond}0.953 &  \cellcolor{tabsecond}0.059\\
        Lego      & \cellcolor{tabfirst}29.65 &  \cellcolor{tabfirst}0.901 & \cellcolor{tabfirst}0.068& \cellcolor{tabsecond}22.88& 0.834 & 0.119 & 22.49 & \cellcolor{tabsecond}0.868& \cellcolor{tabsecond}0.090\\
        Hotdog    & \cellcolor{tabfirst}29.78 &  \cellcolor{tabfirst}0.943& \cellcolor{tabfirst}0.080 & 21.51 & 0.885 & 0.140& \cellcolor{tabsecond}24.59& \cellcolor{tabsecond}0.916& \cellcolor{tabsecond}0.088\\
        Ficus     & \cellcolor{tabfirst}29.47 &  \cellcolor{tabfirst}0.965& \cellcolor{tabsecond}0.052& 23.63 & 0.866 & 0.095& 	\cellcolor{tabsecond}27.23 & \cellcolor{tabsecond}0.964 &  \cellcolor{tabfirst}0.036\\ 
        \midrule
        \textsc{TensoIR}  & \cellcolor{tabfirst}30.96& \cellcolor{tabfirst}0.948& \cellcolor{tabfirst}0.057& 23.92 & 0.873 &0.109	& \cellcolor{tabsecond}26.27& \cellcolor{tabsecond}0.925& \cellcolor{tabsecond}0.068\\
        \textsc{Objaverse}  & \cellcolor{tabfirst}31.77& \cellcolor{tabfirst}0.945& \cellcolor{tabfirst}0.047& 20.64 & 0.825 & 0.171	& \cellcolor{tabsecond}26.44& \cellcolor{tabsecond}0.907& \cellcolor{tabsecond}0.089\\

                \midrule

        \textit{Recons. time}  & \multicolumn{3}{c|}{ \cellcolor{tabfirst}7 s } & \multicolumn{3}{c|}{$\sim$50 min}	& \multicolumn{3}{c}{$\sim$50 min} \\
        \bottomrule
    \end{tabular}
    }
\end{table}



%% file: tables/tbl_exp_ab1.tex
\begin{table}[t]
     \fontsize{8pt}{10pt}\selectfont
    \centering
    \caption{ \textbf{Ablation on \method 's main components}. We compare relighting qualities on \textsc{Objaverse} dataset by removing the fine-grained Gaussians, IDM priors, and light-independent regularization, respectively, showing the effectiveness of our proposed modules. All variants in this table are trained on a lower resolution (256$\times$256).
    The best results are marked in \colorbox{tabfirst}{red}.
    }
    \begin{tabular}{l|cccc}
    \toprule
        Ablations  &   PSNR & SSIM & LPIPS   \\
      \midrule
      Our full model & \cellcolor{tabfirst}{32.44}  & \cellcolor{tabfirst}{0.957} & \cellcolor{tabfirst}{0.043}   \\
      w/o fine-grained Gaussians & 31.46  & 0.945 & 0.047   \\
      w/o IDM priors & 31.21  & 0.947 & 0.049   \\
      w/o light-independence regularization  & 32.07  & 0.954 & 0.044   \\
      w/o neural appearance modeling & 31.72 & 0.950 & 0.047 \\
    \bottomrule
    \end{tabular}
    \label{tab:ablation}
\end{table}

%% file: sec-sig/5_discussion.tex
\subsection{Discussion and Limitations}



\paragraph{Material diversity.}
With our neural material representation, \method can represent diverse types of complex appearances, including furry objects and sub-surface scattering effects. However, there are still types of more complex appearances in the real world, such as highly-reflective or transparent materials, for which \method may suffer from a decreased performance, as we demonstrated in our supplementary.
Dealing with such appearances is orthogonal to our designs, and we leave them for future work. 

\paragraph{GPU memory.}
Like other LRMs, per-pixel Gaussian generation incurs high memory overhead as the number of input views and/or the resolution increase, limiting its scalability. To balance performance and computational cost for practical applications, we use 6 input views in all experiments.
We also investigate the impact of number of input views in our supplementary.

\section{Conclusion}
In this paper, we have proposed \method, a feed-forward relighting framework that directly generates relightable 3D assets from sparse-view inputs by leveraging large model priors, including an LRM and an IDM. Our approach offers flexible relighting capabilities and seamless integration with common rendering pipelines, enabling efficient reproduction of reconstructed 3D assets. Crucially, \method requires no re-training or fine-tuning of large models, relying on relatively small training datasets and affordable computing resources instead. Extensive comparisons demonstrate that \method achieves superior ($\sim$+2.0 dB) relighting quality than previous LRM-based models, and also significantly prevails in rendering speed (by $\sim$25$\times$).

Looking ahead, several promising research directions remain. For instance, integrating large language or multi-modal models to guide material reconstruction via semantic prompts is an interesting avenue. Another valuable addition would be to extend from object-level to scene-level, where multiple assets, global illumination, and spatial contexts can jointly contribute to the final relighting problem, enabling more realistic and coherent relighting results in complex environments. 


%% file: sec-sig/sig_figure_only_pages.tex
\mycfigure{vs_lrm_syn}{vs_lrm_syn.pdf}{
\textbf{Comparison of relighting quality with RelitLRM~\cite{zhang2024relitlrm} on synthetic datasets.}
Our model provides a more reasonable decomposition, resulting in closer alignment with the ground truth across all the test scenes, particularly for complex appearances like fur and sub-surface scattering effects. In contrast, RelitLRM produces inaccurate lighting effects or biased colors.
}

\mycfigure{vs_lrm_real}{vs_lrm_real.pdf}{
\textbf{Comparison of relighting quality with RelitLRM~\cite{zhang2024relitlrm} on real datasets.} Our model generally produces results that are closer to the ground truth, while RelitLRM shows over-blurred textures or biased colors.
}

\mycfigure{vs_image_based}{vs_img_based.pdf}{
\textbf{Comparison of relighting quality with Neural Gaffer~\cite{jin2024neural} and DiLightNet~\cite{zeng2024dilightnet} on synthetic datasets.} \method uses six input views to reconstruct the relightable 3D assets, while other methods directly relight the target view with the target light condition. With the decomposition and underlying 3D representation in \method, our method produces plausible relighting results and a closer match to the ground truth. 
}

\clearpage

%% file: sec-sig/X_suppl.tex
\clearpage
\setcounter{page}{1}
\settopmatter{printacmref=false, printccs=false, printfolios=false} 
\renewcommand\footnotetextcopyrightpermission[1]{} 
\title{Supplementary Material: \\ \method: Feed-Forward Relightable Neural Gaussians}
\makeatletter
\def\@teaserfigures{}
\def\@abstract{}
\def\@keywords{}
\makeatother
\maketitle

We propose \method, a feed-forward relighting framework that leverages large model priors to predict relightable 3D assets from sparse input images. In this supplementary material, we first introduce the algorithm for the saliency map computation in Sec.~\ref{sec:saliency}, then we provide implementation details about the data curation and model training in Sec.~\ref{sec:implementation}. Finally, we provide extra quality validation and metrics in Sec.~\ref{sec:supp_validation}, and discuss some choices of \method in Sec.~\ref{sec:supp_discussion}.

\section{Saliency map computation}
\label{sec:saliency}

In this section, we introduce the algorithm for saliency map computation.
The target of this saliency map is to highlight intrinsic textures, while suppressing boundary effects. First, we measure saliency using the local variance $S_\mathrm{raw}$ from grayscale input images $I_\mathrm{gray}$:
\begin{align}
 \mu &= \mathrm{BoxFilter}(I_\mathrm{gray}, \beta),\\
    \mu_\mathrm{square} & = \mathrm{BoxFilter}(I_\mathrm{gray}^2, \beta) ,  \\
    S_\mathrm{raw} &= \sqrt{\mathrm{max}(\mu_\mathrm{square} - \mu^2, 0)}, 
\end{align}
where $\beta$ is the box filter kernel. 
To suppress the high saliency values that naturally occur around silhouettes, we implement a boundary suppression buffer that attenuates signals near the edges, thereby focusing the refinement on intrinsic surface details. Specifically, we compute the shortest Euclidean distance $D$ from each foreground pixel to the background boundary. This distance is then normalized and clamped to derive a linear weight mask $W_\mathrm{linear}$. Then, a Gaussian kernel is applied to $W_\mathrm{linear}$ to produce a spatially smooth weight map $W$, ensuring a seamless transition between the suppressed boundaries and the preserved internal regions. Finally, the refined saliency map $\mathbf{S}_\mathrm{final}$ is generated by fusing the raw texture details $\mathbf{S}_\mathrm{raw}$ with the boundary weight map $W$:
\begin{align}
    W_\mathrm{linear} & = \mathrm{Clamp}(\mathrm{Normalize}(D), 0, 1),   \\
    W &= \mathrm{GaussianBlur}(W_\mathrm{linear}, \sigma), \\
    \mathbf{S}_\mathrm{final} &= \mathbf{S}_\mathrm{raw} \cdot W.
\end{align}
In practice, we select top-2\% salient patches from each view for geometry synthesis.

\section{Implementation}
\label{sec:implementation}

\subsection{Data curation}
\label{sec:data}

 We sample a subset of the \textsc{Objaverse-xl}~\cite{deitke2023Objaverse} and \textsc{BlenderKit} synthetic assets to build our training set. To ensure high-quality appearance learning, we manually selected a collection of 1000 objects, including a diverse range of materials (45\% diffuse surfaces, 45\% specular materials, and 10\% complex furry textures). 
Each object is rendered from 6 viewpoints, including four orthogonal viewpoints (front, back, left, right) and two oblique top-down views from random directions (45$\degree$ elevation for both, with a 180$\degree$ difference in azimuth), ensuring a reasonable geometric coverage.
For training, each asset is rendered under 100 distinct environment maps, resulting in a comprehensive dataset of 10M images (1000 objects $\times$ 100 views $\times$ 100 environment maps). 
All assets are rendered using the Blender Cycles Engine, with 256 samples per pixel (spp) at 800$\times$800 resolution. 

Compared to RelitLRM~\cite{zhang2024relitlrm}, which is trained on over millions of objects, \method requires only a much smaller dataset (1000 objects) to train. The reason for this difference is their underlying learning paradigms. RelitLRM is a data-driven generative model that relies on massive parameters ($\sim$0.4B) to learn the complex mapping from input 2D images and environment maps to the final radiance fields.  In contrast, \method builds upon RelitLRM's pre-trained latent spaces, and learns only a mapping from RelitLRM's latent space to another relightable latent space, and most importantly, with the guidance of material-specific priors. Consequently, \method achieves superior material decomposition from only 1000 objects, which is more affordable and easily accessible by common users.

\subsection{Model and training details}
\label{sec:model}

The MaterialFormer is a 4-layer Transformer with a hidden dimension of 1024. We use Pre-Layer Normalization (Pre-LN) and GeLU activation functions to ensure stable training and faster convergence. 
Following RelitLRM, we keep the $8 \times 8$ patch size for all input 2D images and IDM priors. When obtaining IDM priors from DiffusionRenderer~\cite{liang2025diffusion}, we organize the sparse-view inputs into a video stream to ensure cross-view consistency. 
We employ the $\mathcal{L}_2$ loss and perceptual loss~\cite{chen2017photographic} to the final renderings. The full loss function can be formulated as follows:

\begin{align}
   \mathcal{L} =& \lambda_\mathrm{1}\mathcal{L}_{2} + \lambda_\mathrm{per} \mathcal{L}_\mathrm{per} + \mathcal{L}_\mathrm{ind}.
\end{align}
In practice, $(\lambda_c, \lambda_\mathrm{KLD}, \lambda_1, \lambda_{per} ) = (0.2, 0.2, 1.0, 0.5)$.
The model is trained at $512 \times 512$ resolution and converges in 9 days on 8 NVIDIA L40 (48GB) GPUs.

\subsection{Universal neural appearance decoder}

The universal neural appearance decoder is a 4-layer 256-unit MLP, and we do not perform the shading frame conversion, since the shading normal at Gaussians can be difficult to define, especially for fluffy objects. 
We perform a deferred shading pipeline to render our Gaussian assets under environment lights. First, we blend the neural features from each Gaussian along the camera ray to get an aggregated neural feature for each pixel in the image space, then we decode it with our universal neural appearance decoder into reflectance. To render under environment lights, we importance sample the environment light with 64 spp during training and 128 spp during inference, and only 24 out of 64 samples are dedicated to gradient computation for acceleration.

\section{Additional validation}
\label{sec:supp_validation} 


\subsection{More visual comparisons with RelitLRM}
\begin{figure*}[p]
    \centering
    \includegraphics[
        width=\textwidth,
        height=\textheight,
        keepaspectratio
    ]{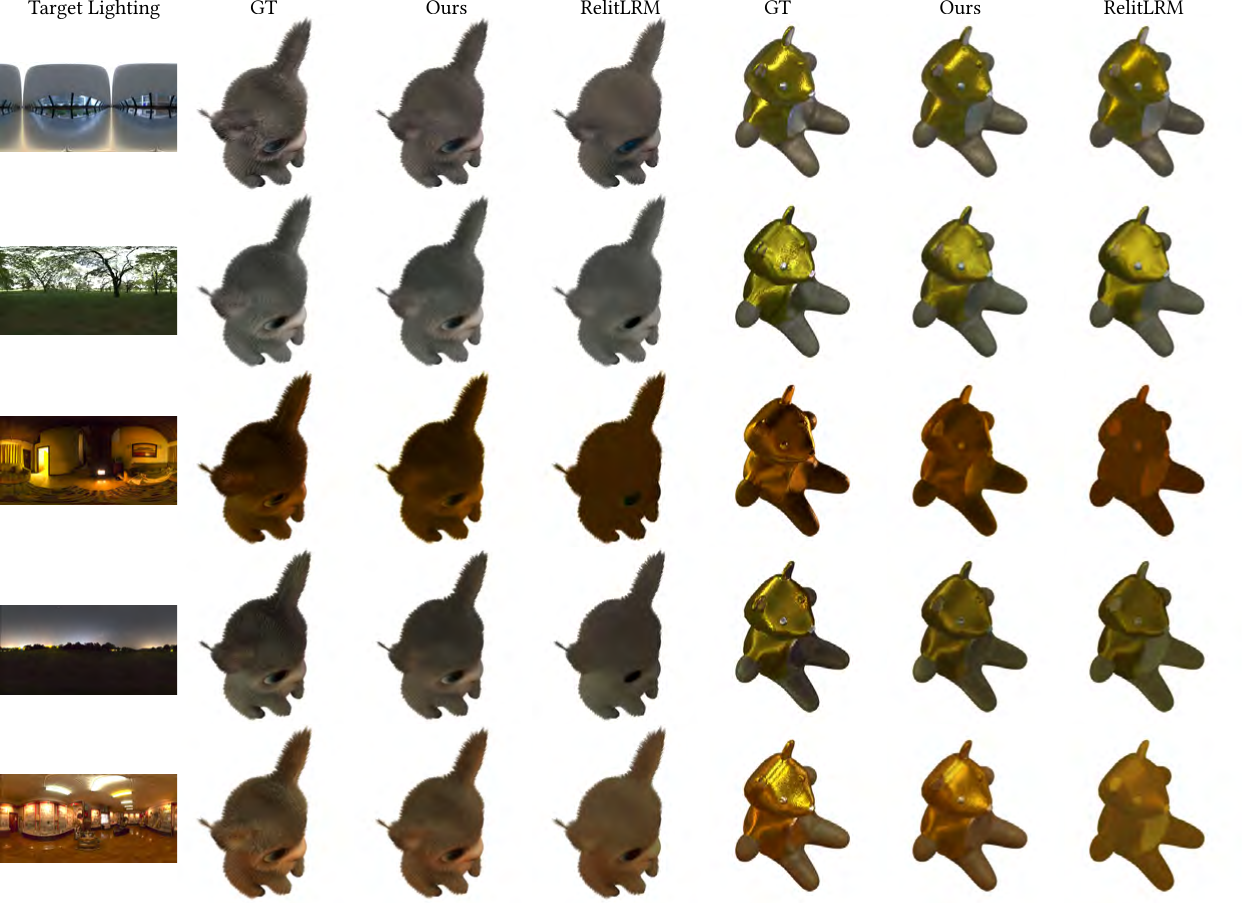}
    \caption{\textbf{Comparison of the relighting quality with RelitLRM on synthetic datasets.}
Our model provides more reasonable decomposition results, especially for complex appearances (e.g., fur).}
    \label{fig:vs_lrm_more}
\end{figure*}

\begin{figure*}[p]
    \centering
    \includegraphics[
        width=\textwidth,
        height=\textheight,
        keepaspectratio
    ]{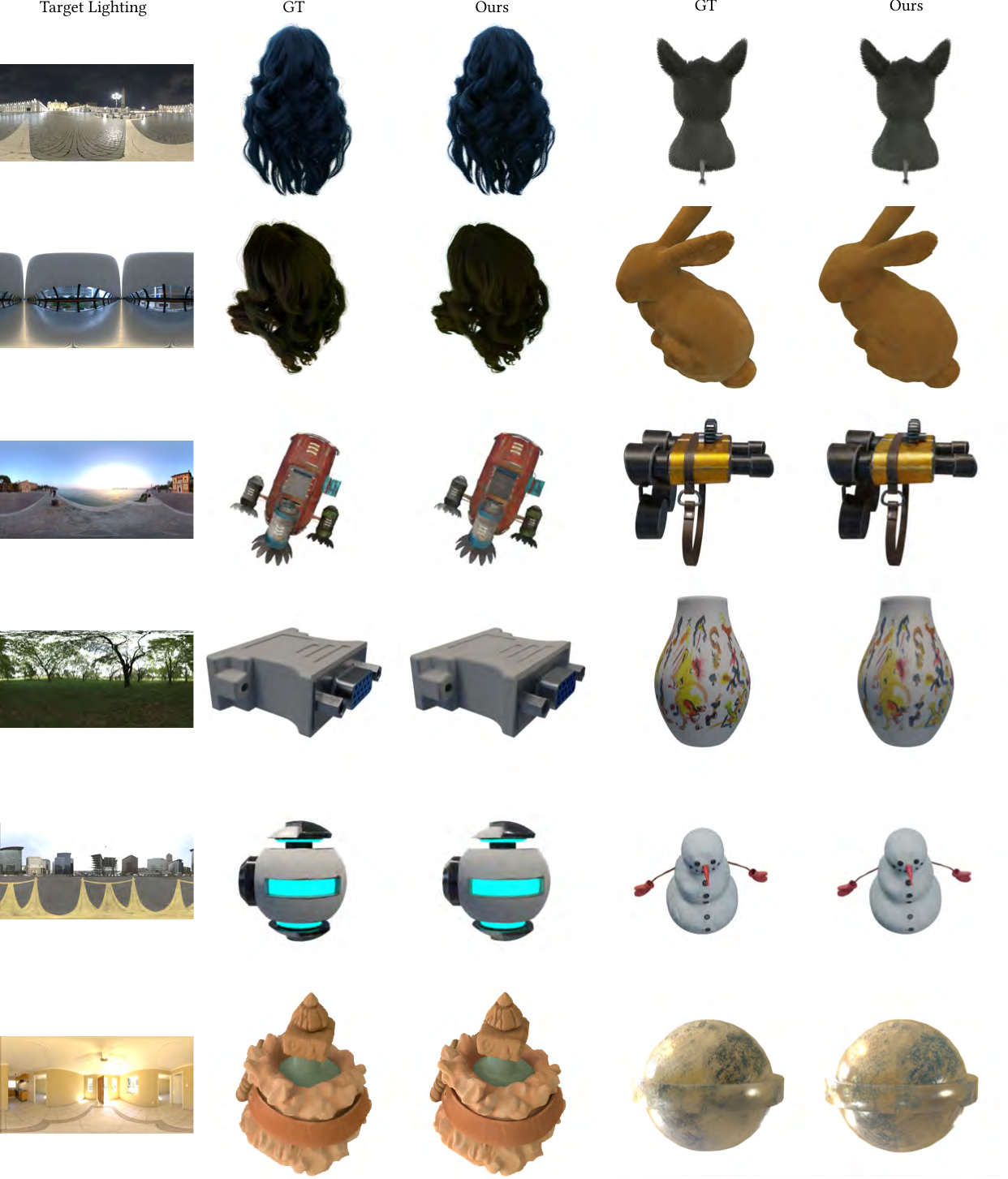}
    \caption{\textbf{Relighting results of \method under various environment maps.}
Our model can plausibly decompose the light and material, leading to high-quality relighting results under varying environment lights.}
    \label{fig:gallery}
\end{figure*}

In the main paper, we compared \method with RelitLRM~\cite{zhang2024relitlrm} regarding the relighting quality. In Fig.~\ref{fig:vs_lrm_more}, we present additional results of multiple objects under various environmental lights. \method consistently produces renderings that are closer to the ground truth across different materials, showing more detailed appearances for both surface-like and furry-like materials. 
Some extra visualizations of our results are also exhibited in Fig.~\ref{fig:gallery}.

\subsection{Visual comparisons with dense-view 3DGS methods}

\myfigure{vs_optim_syn}{vs_optim_syn.pdf}{
\textbf{Comparison of relighting quality with dense-view 3DGS-based relighting methods~\cite{liang_2024_gsir, gao_2023_relightablegs} on synthetic datasets.} Our model produces the best quality with only 6 input views, compared to more than 100 input images from other methods. 
}

In the main paper, we provided the quantitative metrics of our relighting quality comparison with dense-view optimization-based 3DGS methods~\cite{gao_2023_relightablegs, liang_2024_gsir}. In Fig.~\ref{fig:vs_optim_syn}, we provide the visual comparisons under various lighting conditions. \method consistently shows less artifacts and closer matches to the ground truth, while using far fewer input images (6 views) compared to the dense inputs ($>$100 views) required by those optimization-based approaches.

\subsection{Comparison of NVS quality}

\myfigure{nvs}{nvs.pdf}{
\textbf{Comparison of NVS qualities with DNG~\cite{li2024dngaussian} on synthetic datasets.}
With both 6 input views, our approach achieves superior results across diverse scenes than DNG.
}

In Fig.~\ref{fig:nvs} and Table~\ref{tab:tensoir_comp1}, we compare the NVS quality with a sparse-view optimization-based 3DGS approach, DNG~\cite{li2024dngaussian}, on a synthetic dataset. With both methods utilizing only six input images, our approach achieves higher quality in most cases and on average ($\sim$+1.5 dB). Furthermore, our reconstruction is $\sim$17$\times$ faster than DNG, thanks to its feed-forward design.


\subsection{NVS Quality and number of input views}

Regarding the effect of input views, we also provide a quantitative comparison on NVS qualities, including \method and representative optimization-based 3DGS methods using varying numbers of input views in Table~\ref{tab:num_input}. Note that, like other LRM-based approaches, \method is limited to sparse inputs due to the per-pixel Gaussian paradigm, which easily incurs prohibitive GPU memory costs when the number of input views increases. Despite this, \method still achieves comparable or superior quality with only six input views, against these optimization-based methods using 25 or more views. Furthermore, \method also enjoys a much faster reconstruction time.



\begin{table}[htb]
    \centering
    \caption{
    \textbf{Comparison of our NVS quality with sparse-view 3DGS reconstruction methods~\cite{li2024dngaussian}. } 6 input views are used for each scene. \method shows an overall better performance with significantly higher speed in reconstruction. The best results are marked in \colorbox{tabfirst}{red}. Visual comparisons are provided in Fig.~\ref{fig:nvs}.
    }
    \label{tab:tensoir_comp1}
    \vspace{-7pt} 
    {
    \begin{tabular}{l|ccc|ccc}
        \toprule
        Scene & \multicolumn{3}{c|}{Ours}  & \multicolumn{3}{c}{DNG~\cite{li2024dngaussian}} \\
        & PSNR & SSIM & LPIPS & PSNR & SSIM & LPIPS  \\ 
        \midrule
        Armadillo & \cellcolor{tabfirst}{36.67}&\cellcolor{tabfirst}{0.987} &\cellcolor{tabfirst}{0.028}
& 28.88 & 0.953 & 0.049 \\
        Lego      & \cellcolor{tabfirst}{30.51}& \cellcolor{tabfirst}{0.921} & \cellcolor{tabfirst}{0.056}
& 26.09 & 0.878 & 0.119 \\
        Hair    & \cellcolor{tabfirst}{26.34} & \cellcolor{tabfirst}{0.903} & \cellcolor{tabfirst}{0.074}
& 23.54 & 0.856 & 0.127 \\
        Knight     & \cellcolor{tabfirst}{31.34} &\cellcolor{tabfirst}{0.978} & 0.054 & 29.41 & 0.956 & \cellcolor{tabfirst}{0.049} \\ 
        \midrule
        \textit{Average}  & \cellcolor{tabfirst}{30.62} &\cellcolor{tabfirst}{0.950}& \cellcolor{tabfirst}{0.059} & 26.98 & 0.911 & 0.086 \\
                \midrule
        \textit{Recons. time}  & \multicolumn{3}{c|}{ 7 s } & \multicolumn{3}{c}{$\sim$2 min} \\
        \bottomrule
    \end{tabular}
    }
\end{table}

\begin{table}
    \centering
    \caption{
    \textbf{
   Comparison of NVS quality with 3DGS-based methods using varying input views on the Lego scene from TensoIR~\cite{jin2023tensoir}.} Our method, \method, achieves comparable quality with significantly fewer input views and provides substantially faster reconstruction than all optimization-based baselines. All provided values are PSNR. The last column shows the reconstruction time for all methods using 6 input views.
    }
    \vspace{-7pt} 
    \begin{tabular}{l|cccc|c}
    \toprule
         \#Input views &  100&  50&  25&  6 & \textit{Recons. time}\\
     \midrule
         Ours&  -&  -&  -&   30.51 & 7 s\\
         R3DG&  34.12 & 32.04 & 26.85  & 19.32 &  $\sim$40 min\\
         DNG&  30.65 & 28.32 & 26.43 & 26.09 &  $\sim$2 min\\
         3DGS& 20.33 & 11.34 & 7.98  & 6.45  & $\sim$10 min\\
     \bottomrule
    \end{tabular}
\label{tab:num_input}
\end{table}



\begin{table}[htb]
    \centering
    \caption{\textbf{Comparison with image-based relighting models on \textsc{Objects-with-Lighting} Dataset~\cite{Ummenhofer2024OWL}.} While LightSwitch~\cite{litman2025lightswitch} achieves higher PSNR and SSIM, it suffers from significantly slower relighting speed (much slower than ours) and inferior visual quality, as shown in Fig.~\ref{fig:vs_ls_owl}. The best/second-best results are marked in \colorbox{tabfirst}{red}/\colorbox{tabsecond}{yellow}.}
    \label{tab:owl_3d_comparison}
    \vspace{-7pt} 
    \begin{tabular}{l|ccc|c}
    \toprule
    Method &  {PSNR } & {SSIM } & {LPIPS} & \textit{Relighting time}  \\
 \midrule
    Ours & \cellcolor{tabfirst}{26.03}   &  \cellcolor{tabfirst}{0.841}    &  \cellcolor{tabfirst}{0.244}   & \cellcolor{tabfirst}{0.3 s}   \\ 
     LightSwitch & \cellcolor{tabsecond}{25.43}   & \cellcolor{tabsecond}{0.840}    & \cellcolor{tabsecond}{0.297}  & 60 s \\
    DiffusionRenderer & 21.57  & 0.795    & 0.341  & \cellcolor{tabsecond}{28 s}    \\ 
    \bottomrule
    \end{tabular}
    
\end{table}



\begin{table}[htb]
    \centering
    \caption{
    \textbf{The impact of number of input views. } We evaluate the model with different number of input views on \textsc{Objaverse}~\cite{deitke2023Objaverse} and \textsc{TensoIR}~\cite{jin_2023_tensoir} datasets.
    We choose to use 6 input views (marked as \textbf{bold}) in our experiments.
    Visual comparisons are provided in Fig.~\ref{fig:inputview}.
    }
    \label{tab:view}
    \vspace{-7pt} 
    {
    \begin{tabular}{l|ccc|ccc}
        \toprule
        Input views & \multicolumn{3}{c|}{\textsc{Objaverse}}  & \multicolumn{3}{c}{\textsc{TensoIR}} \\
        & PSNR & SSIM & LPIPS & PSNR & SSIM & LPIPS  \\ 
        \midrule
        2 & 24.84 & 0.901 & 0.103 & 24.53 & 0.889 & 0.114 \\
        4 & 28.31 & 0.916 & 0.069 & 27.75 & 0.908 & 0.077 \\
        \textbf{6} & \textbf{31.77} & \textbf{0.945} & \textbf{0.047} & \textbf{30.96} & \textbf{0.948} & \textbf{0.057} \\
        8 & 31.83 & 0.950 & 0.049 & 31.14 & 0.948 & 0.057 \\
        10 & 31.95 & 0.951 & 0.049 & 31.28 & 0.949 & 0.054 \\
        \bottomrule
    \end{tabular}
    }
\end{table}

\begin{table}[htb]
    \centering
    \caption{\textbf{The impact of $K$.} In \method, we select top-$K$ salient patches to perform the geometry synthesis, and we test the effect of different $K$ choices on \textsc{Objaverse} dataset. We choose $K=2\%$ in our paper (marked as \textbf{bold}).}
    \label{tab:k}
    \vspace{-7pt} 
    \begin{tabular}{l|ccc}
    \toprule
    $K$ &  {PSNR } & {SSIM } & {LPIPS} \\
 \midrule
    1\% & 31.05 & 0.940 & 0.052 \\
    \textbf{2\%} & \textbf{31.77} & \textbf{0.945} & \textbf{0.047} \\ 
    4\% & 31.70 & 0.945 & 0.046 \\ 
    8\% & 31.54 & 0.943 & 0.047 \\ 

    \bottomrule
    \end{tabular}
    
\end{table}

\subsection{Comparison with image-based relighting models}

\begin{figure*}[p]
    \centering
    \includegraphics[
        width=0.9\textwidth,
        height=\textheight,
        keepaspectratio
    ]{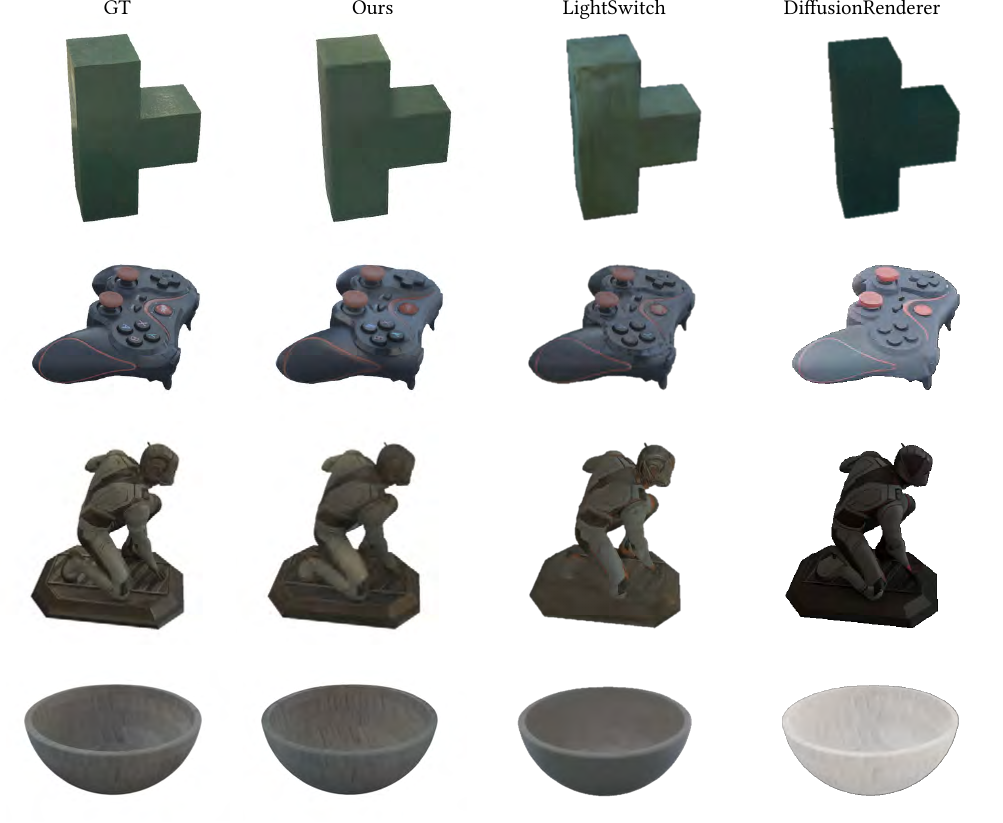}
    \caption{\textbf{Comparison with image-based relighting methods (DiffusionRenderer~\cite{liang2025diffusion}, LightSwitch~\cite{litman2025lightswitch}) on real-world datasets.} 
    Our method produces overall closer results to the ground truth.
    }
    \label{fig:vs_ls_owl}
\end{figure*}

In the main paper, we have compared \method with Neural Gaffer~\cite{jin2024neural} and DiLightNet~\cite{zeng2024dilightnet} on synthetic datasets. In this section, we compare with LightSwitch~\cite{litman2025lightswitch} and DiffusionRenderer~\cite{liang2025diffusion} on real-world datasets. 
For DiffusionRenderer, we encode the input views as video sequences to perform the relighting.
For Light\-Switch~\cite{litman2025lightswitch}, since we failed to re-produce the results reported in their paper by the released codes, we only compare on the \textsc{Owl}~\cite{Ummenhofer2024OWL} dataset using the images provided directly by the authors.
Visual comparison and quantitative metrics are provided in Fig.~\ref{fig:vs_ls_owl} and Table~\ref{tab:owl_3d_comparison}. Our method achieves both higher metrics and visual qualities than both methods. 
Apart from our significantly faster relighting speed, LightSwitch also suffers from much slower reconstruction time ($\sim$40$\times$) than \method, as it optimizes the geometry during reconstruction.

\subsection{Robustness against input light conditions}

\begin{figure*}[p]
    \centering
    \includegraphics[
        width=\textwidth,
        height=\textheight,
        keepaspectratio
    ]{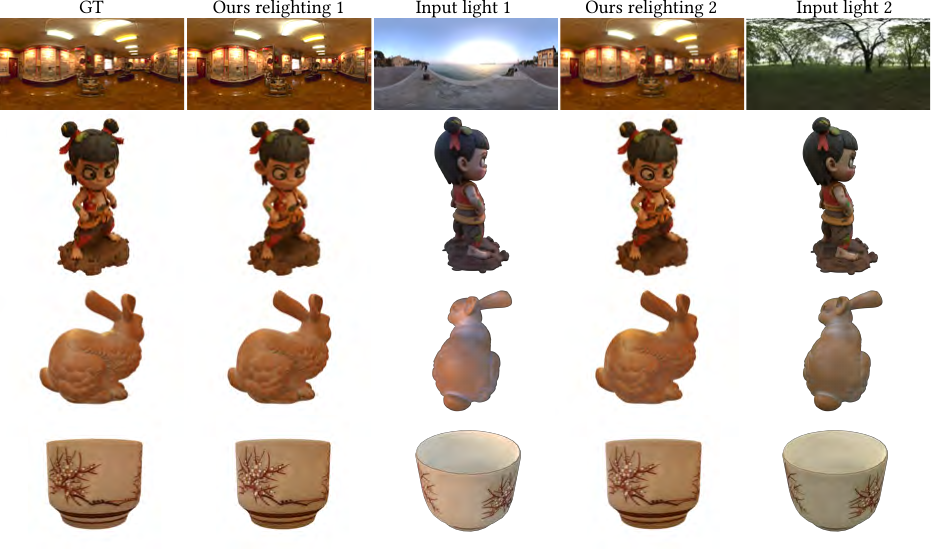}
    \caption{
    \textbf{Robustness of our decomposition against different input light conditions.} We validate our relighting results of the same objects with different initial light conditions, and both predicts close results to the ground truth, demonstrating the plausible decomposition in \method.
}
    \label{fig:varying_inputs}
\end{figure*}

By the proposed MaterialFormer and light-independence regularization, we decompose the light and materials from input radiance fields to achieve high-fidelity relighting. In Fig.~\ref{fig:varying_inputs}, we demonstrate the robustness of our decomposition against different input light conditions. By feeding the same object with different initial light conditions to our model, \method predicts close results to the ground truth under both conditions, showing the plausible decomposition.

\section{Additional discussion}
\label{sec:supp_discussion}

\begin{figure}[p]
    \centering
    \includegraphics[
        width=0.42\textwidth,
        height=0.42\textheight,
        keepaspectratio
    ]{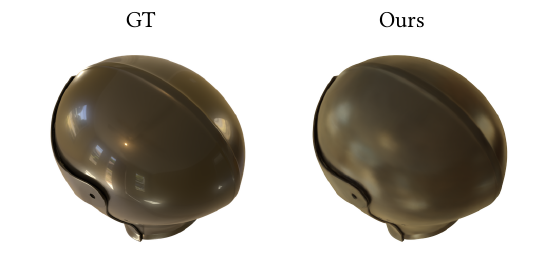}
    \caption{\textbf{Relighting highly-reflective objects.} Our method tends to produce blurry results when rendering highly-reflective materials.
}
    \label{fig:fail}
\end{figure}

\paragraph{Highly-reflective objects.}
As illustrated in Fig.~\ref{fig:fail}, the performance of our model drops with highly specular materials, resulting in blurred reflections. Dealing with such appearances is orthogonal to our designs, and we leave them for future work.

\myfigure{ablation_loss}{heatmap.pdf}{
\textbf{The effect of light-independence regularization.}
With the light-independence regularization, the MaterialFormer can predict a more compact and plausible latent distribution for the same material. Note that larger values indicate higher similarity. 
}

\paragraph{The effect of light-independence regularization.}
In Fig.~\ref{fig:ablation_loss}, we show the effect of this regularization by analyzing the numerical distribution of the learned latent space. With such regularization, the output from MaterialFormer is forced to be light-independent.

\begin{figure*}[p]
    \centering
    \includegraphics[
        width=\textwidth,
        height=\textheight,
        keepaspectratio
    ]{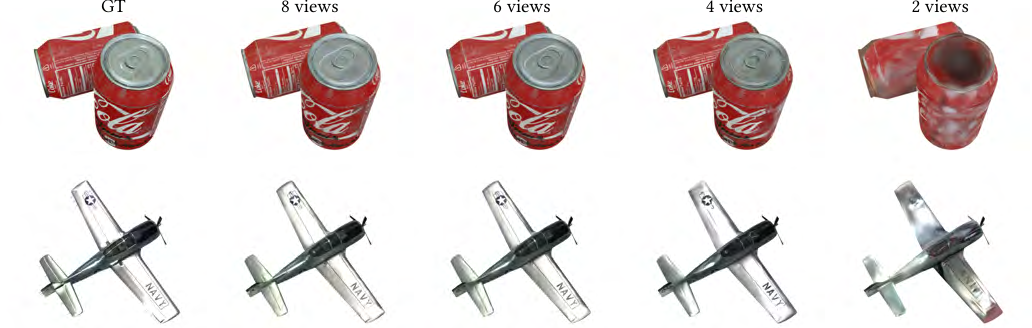}
    \caption{\textbf{Impact of the number of Input views.} 
    Although \method supports varying number of input views, to balance the quality and cost, we choose to use 6 input views for all experiments in our paper.
}
    \label{fig:inputview}
\end{figure*}

\paragraph{Impact of input view numbers.} 
We analyze the influence of the number of input views in Fig.~\ref{fig:inputview} and Table~\ref{tab:view}. Increasing the number of input views consistently improves rendering quality, while with more than 6 views, the improvement becomes marginal. Therefore, we choose to use 6 input views to balance quality and efficiency. 

\paragraph{Impact of $K$ in salient patch selection.}
We choose top-$K$ patches from each input view to perform the fine-grained geometry synthesis. In Table~\ref{tab:k}, we show the relighting quality with different $K$ values. 
Larger $K$ indicates more newly generated fine-grained Gaussians, which is intuitively a benefit. However, too much redundant Gaussians can still break the original geometry structure generated from the prior LRM, and also makes it difficult for the model to converge.

\clearpage

